\documentclass[ aps,showpacs,jcp,superscriptaddress,floatfix,twocolumn,citeautoscript
]{revtex4-1}
\usepackage{url}
\usepackage{hyperref}
\usepackage{amsmath}
\usepackage{amssymb}
\usepackage[usenames]{color}
\usepackage{graphicx}
\usepackage{dcolumn}
\usepackage{bm}
\usepackage{booktabs}
\usepackage{multirow}
\hypersetup{pdfborder=0 0 0,colorlinks=true,citecolor=blue,linkcolor=blue}
\begin{document}
\title{Phononic thermal conductivity in silicene: the role of vacancy defects and boundary scattering}

\author{M. Barati}
\affiliation{Department of Physics, Shahid Beheshti University, G. C., Evin, Tehran 1983969411, Iran}
\author{T. Vazifehshenas}%
\email{t-vazifeh@sbu.ac.ir}
\affiliation{Department of Physics, Shahid Beheshti University, G. C., Evin, Tehran 1983969411, Iran}
 \author{T. Salavati-fard}
 \affiliation{Department of Physics and Astronomy, University of Delaware, Newark, DE 19716, USA}
 \author{M. Farmanbar}
\affiliation{Faculty of Science and Technology and MESA+ Institute for Nanotechnology, University of Twente, P.O. Box 217, 7500 AE Enschede, The Netherlands}


\begin{abstract}
We calculate the thermal conductivity of free-standing silicene using the phonon Boltzmann transport equation within the relaxation time approximation. In this calculation, we investigate the effects of sample size and different scattering mechanisms such as phonon-phonon, phonon-boundary, phonon-isotope and phonon-vacancy defect. Moreover, the role of different phonon modes is examined. We show that, in contrast to graphene, the dominant contribution to the thermal conductivity of silicene originates from the in-plane acoustic branches, which is about 70\% at room temperature and this contribution becomes larger by considering vacancy defects. Our results indicate that while the thermal conductivity of silicene is significantly suppressed by the vacancy defects, the effect of isotopes on the phononic transport is small. Our calculations demonstrate that by removing only one of every 400 silicon atoms, a substantial reduction of about 58\% in thermal conductivity is achieved. Furthermore, we find that the phonon-boundary scattering is important in defectless and small-size silicene samples, specially at low temperatures.    
\end{abstract}

\maketitle
\section{\label{sec:level1}Introduction}
Interest in two-dimensional (2D) materials and layered structures has been considerably growing during last two decades, because of a wide variety of applications in nanotechnology. The discovery of graphene\cite{key-1}, a planar honeycomb structure of carbon atoms, was a milestone in the development of nano sciences due to its extraordinary properties such as high carrier mobility \cite{key-1,key-2,key-3,key-4,key-5}, room-temperature ballistic carrier transport \cite{key-1,key-3}, high lattice stability \cite{key-2,key-3} and high mechanical strength \cite{key-3}.

\begin{figure}
 \includegraphics[width=7cm]{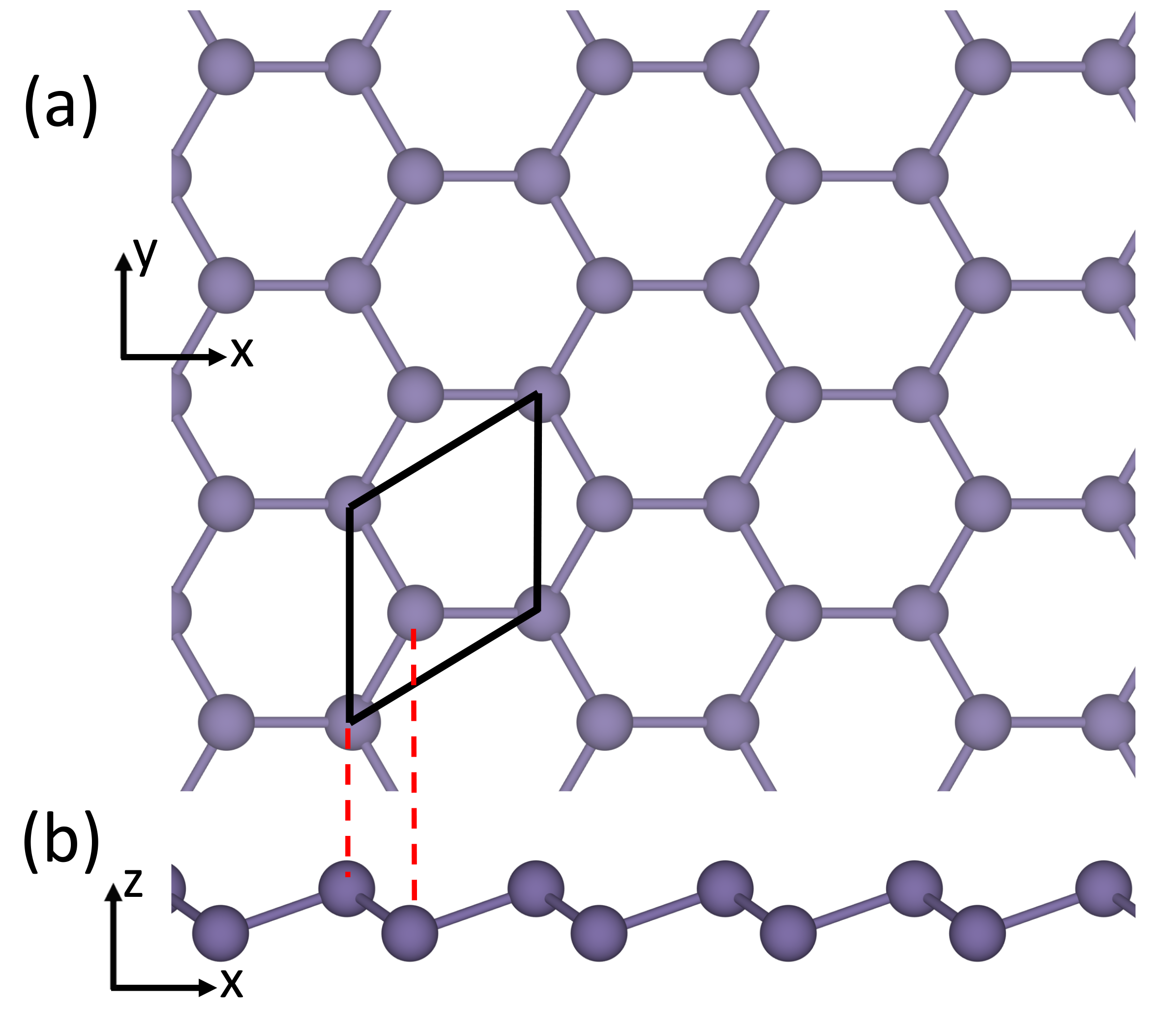} 
 \caption{(a) Top  and (b) side views of silicene. The unit cell is shown in  black lines.} \label{fig1:sil}
\end{figure}

Silicene, the 2D graphene-like structure of silicon, is slightly buckled (Fig. \ref{fig1:sil}) in its most stable form. The buckling leads to some new characteristics \cite{key-6,key-7} which make silicene a promising  alternative to graphene in the rapidly developing fields of thermoelectrics and nanoelectronics \cite{key-8,key-9}. While the electronic properties of silicene have been extensively studied \cite{key-10,key-11}, its thermal transport features are still open for more investigations \cite{key-9}. Graphene exhibits a high lattice thermal conductivity \cite{key-12,key-13,key-14} which may be useful in some applications such as electronic cooling and heat spreading \cite{key-15,key-16}. Nevertheless, high lattice thermal conductivity is not appealing in thermoelectric devices where a very low thermal conductivity is required. Unlike graphene, silicene with a small lattice thermal conductivity \cite{key-9} can be effectively used in thermoelectricity in future.
Because of difficulties in synthesizing free-standing silicene, the thermal conductivity of silicene has not been experimentally measured yet, but it has been calculated using a couple of numerical methods. Molecular dynamics (MD) simulations, which rigorously rely on the empirical interatomic potentials, predicted values of thermal conductivity in the range of 5 to 69 W/mK \cite{key-17,key-18,key-19,key-20}. On the other hand, the predicted values of thermal conductivity of silicene from first-principle based calculations span 20 to 30 W/mK \cite{key-9,key-21,key-22}. In this approach, the harmonic second-order and anharmonic third-order interatomic force constants \cite{key-23,key-24,key-25,key-26,key-27,key-28,key-29,key-30,key-31} are determined from first-principles using density functional perturbation theory as input parameters and then the phonon Boltzmann transport equation (PBTE) is solved. PBTE ia a successful microscopic description for the intrinsic thermal conductivity of semiconductors and insulators which was first formulated by Peierls \cite{key-32}. In PBTE, the variation of the phonon distribution with the temperature gradient is balanced by its changes due to the scattering processes. Instead of solving the PBTE self-consistently, which is computationally very expensive, it is common to employ the relaxation time approximation (RTA) in which the phonon lifetime is approximated as the sum over all the contributions from different scattering mechanisms \cite{key-33,key-34}. Comparing the results of full iterative calculations of intrinsic lattice thermal conductivity of silicene \cite{key-21,key-35} with those obtained from the RTA \cite{key-9}, shows that the RTA works well for the case of silicene.

In this study, we use a theoretical approach to calculate the lattice thermal
conductivity of free-standing monolayer silicene based on a numerical solution of the PBTE within RTA. We use the second- and third-order force constants obtained from the density functional theory (DFT) to calculate the phononic thermal conductivity under different scattering mechanisms including the phonon-phonon, phonon-boundary, phonon-isotope and phonon-vacancy scatterings and study the effects of temperature and size of the system. Also, we investigate how various phonon modes contribute to the thermal conductivity and show that the in-plane acoustic modes carry most of the heat, in contrast to the case of graphene. Furthermore, we find that, among different elastic scattering mechanisms, the phonon-vacancy scattering has the most significant influence on the silicene thermal conductivity. This prediction for the vacancy effect is in good agreement with that of the MD simulation. Our findings shed some light on how the figure of merit and thermoelectricity in silicene can be controlled.  

The rest of the paper is organized as follows: in section II, the theory is given and the method is explained in detail. The results are presented and discussed thoroughly in section III. Finally, the highlights of this work  are summarized in section IV.

\section{\label{sec:level2}Theory and Method}

We consider a free-standing monolayer silicene and calculate the thermal conductivity based on the linearized PBTE within the RTA. As we mentioned earlier, the PBTE can be solved using an iterative method too, however, it has been shown that the difference between calculated thermal conductivity of silicene by employing any of the two methods is small \cite{key-9}. 

\begin{figure}
 \includegraphics[width=8.5cm]{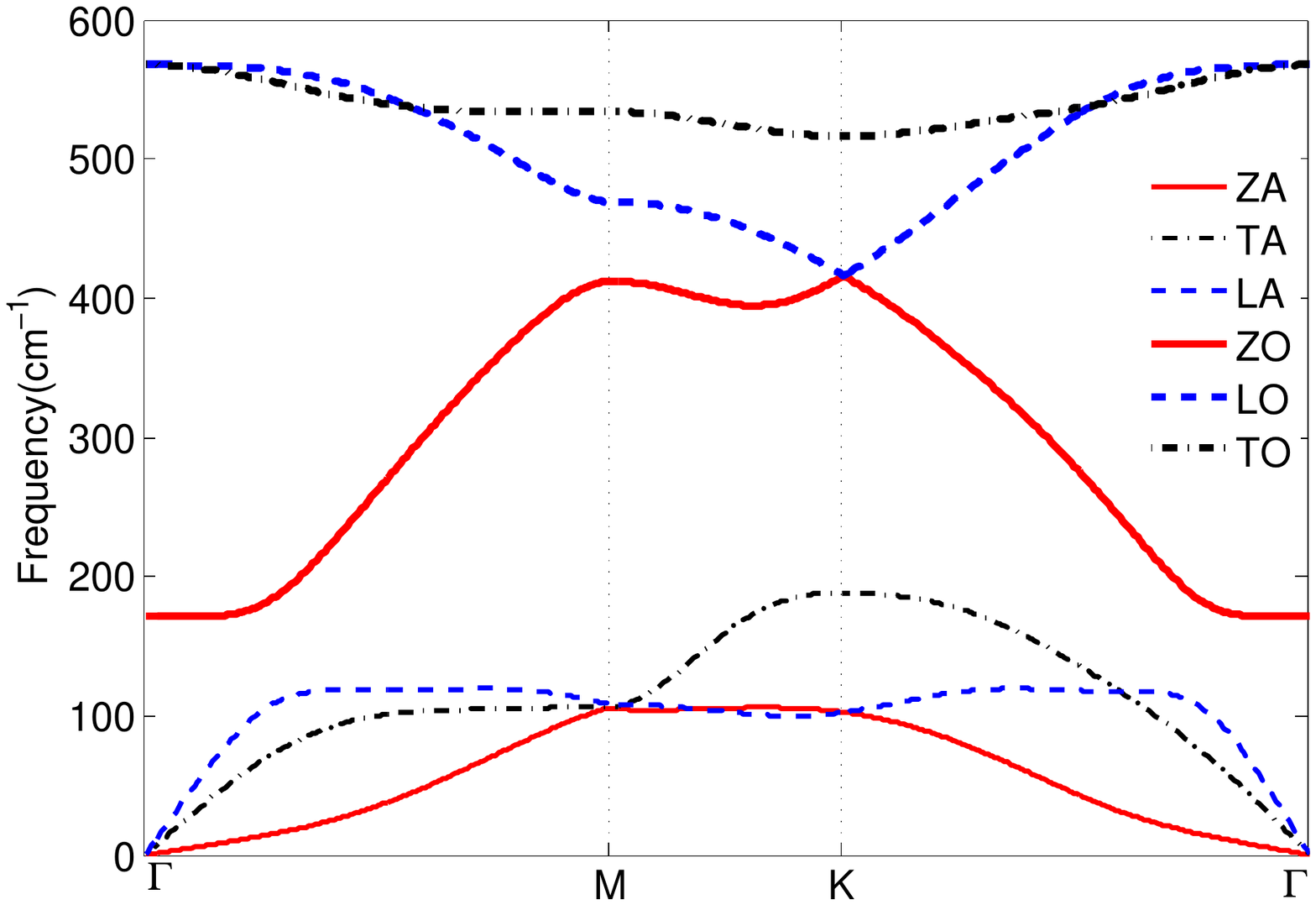} 
 \caption{Phonon dispersion of monolayer silicene.} \label{fig2:sil}
  \end{figure}

The contribution of all phonon modes to the thermal conductivity, $\kappa$, can be expressed as: \cite{key-36,key-37}
\begin{equation}
\kappa=\frac{\hbar^{2}}{Vk_{B}T^{2}}\sum_{\lambda}\omega_{\lambda}^{2}v_{\lambda}^{2}n_{\lambda}^{0}(1+n_{\lambda}^{0})\tau_{\lambda},
\label{eq:1}
\end{equation}
where $v_{\lambda}$, $n_{\lambda}^{0}$ and $\tau_{\lambda}$ are the phonon velocity, equilibrium distribution function and scattering time, respectively. Also, $V$ is the volume of the unit cell. The phonon frequency of mode ${\lambda}$, $\omega_{\lambda}$, is obtained from the dynamical matrix of the system, which its (${\alpha\beta}$) component, $D^{\alpha\beta}(\mathbf{q})$, is given by:
\begin{equation}
D_{jj^{'}}^{\alpha\beta}(\mathbf{q})=\frac{1}{\sqrt{M_{j}M_{j^{'}}}}\sum_{\mathbf{R^{'}}}\phi_{\mathbf{0}j\mathbf{R^{'}}j^{'}}^{\alpha\beta}e^{i\mathbf{q}.\mathbf{R}^{'}},
\label{eq:2}
\end{equation}
Here, $\phi^{\alpha\beta}$ is the harmonic (second-order) force constant which can be extracted from the first-principle calculations, $\mathbf{R}^{'}$ is the lattice vector of unit cell and  $M_{j}$ is the mass of $j$th atom in this unit cell. Since there is only one type of atom in silicene, so we set $M_{j/j^{'}}=M$. The dispersion relations of different branches of phonons can be obtained from the zeros of the determinant of the dynamical matrix. Fig. \ref{fig2:sil} shows the phonon dispersion of free-standing silicene along some high symmetry directions. There are three acoustic and three optical phonon branches as a result of two atoms per unit cell in silicene.
In contrast to graphene, due to buckled structure of silicene, the flexural modes of silicene are not completely out-of-plane and can get coupled to the in-plane modes. As a result of this coupling, the selection rules for the phonon-phonon scattering in silicene is different from graphene.

It is well-known that the scattering processes play an important role in determining the thermal conductivity. Phonon-phonon, phonon-boundary, phonon-isotope and phonon-vacancy scatterings are the main mechanisms for the phonon transport. Within the RTA, the scattering lifetime,  $\tau_{\lambda}$, can be calculated by employing the Matthiessen's rule as \cite{key-33,key-34}: 
\begin{equation}
\frac{1}{\tau_{\lambda}}=\sum_{i}\frac{1}{\tau_{\lambda}^{i}},
\label{eq:3}
\end{equation}
where $\tau_{\lambda}^{i}$ denotes the relaxation time of each scattering mechanism and the sum goes over all the possible mechanisms.

In a perfect lattice without any defects or rough boundaries, thermal conductivity is limited by the phonon-phonon scattering through the absorption and emission processes in which the following energy and quasi-momentum conservation laws are satisfied \cite{key-23}:

\begin{equation}
\omega_{\lambda}\pm\omega_{\lambda^{'}}=\omega_{\lambda^{''}}
\label{eq:4}
\end{equation}
\begin{equation}
\mathbf{q}\pm \mathbf{q}^{'}=\mathbf{q}^{''}+\mathbf{G}, 
\label{eq:5}
\end{equation}
Here, $\mathbf{G}$ is a reciprocal lattice vector. The relaxation time of the phonon-phonon scattering, $\tau_{\lambda}^{Ph}$, can be expressed as \cite{key-29}
\begin{equation}
\tau_{\lambda}^{Ph}=\frac{n_{\lambda}^{0}(1+n_{\lambda}^{0})}{\sum_{\lambda^{'}\lambda^{''}}[W_{\lambda\lambda^{'}\lambda^{''}}^{+}+\frac{1}{2}W_{\lambda\lambda^{'}\lambda^{''}}^{-}]},
\label{eq:6}
\end{equation}
where $W_{\lambda\lambda^{'}\lambda^{''}}^{\pm}$ is the phonon-phonon
scattering rate for the absorption and emission processes \cite{key-38,key-39,key-40}
\begin{eqnarray}
W_{\lambda\lambda^{'}\lambda^{''}}^{\pm} & = & \frac{\pi\hbar}{4N_{0}}\frac{(1+n_{\lambda}^{0})(n_{\lambda^{'}}^{0}+\frac{1}{2}\pm\frac{1}{2})n_{\lambda^{''}}^{0}}{\omega_{\lambda}\omega_{\lambda^{'}}\omega_{\lambda^{''}}}\nonumber\\
 &  & \left|V_{\lambda\lambda^{'}\lambda^{''}}^{\pm}\right|^{2}\delta(\omega_{\lambda}\pm\omega_{\lambda^{'}}-\omega_{\lambda^{''}}).
 \label{eq:7}
\end{eqnarray}
In the above expression $N_{0}$ is the number of unit cells and the three-phonon scattering matrix elements of the phonon-phonon interaction, $V_{\lambda\lambda^{'}\lambda^{''}}^{\pm}$,
are given by:

\begin{eqnarray}
V_{\lambda\lambda^{'}\lambda^{''}}^{\pm} & = & \sum_{j}\sum_{R^{'},j^{'}}\sum_{R^{''},j^{''}}\sum_{\alpha\beta\gamma}\phi_{0j,R^{'}j^{'},R^{''}j^{''}}^{\alpha\beta\gamma}\nonumber \\
 &  & \frac{e_{\lambda}^{\alpha}(j)e_{\pm\lambda^{'}}^{\beta}(j^{'})e_{\lambda^{''}}^{\gamma}(j^{''})}{\sqrt{M_{j}M_{j^{'}}M_{j^{''}}}}e^{i(\pm \mathbf{q}^{'}).\mathbf{R}^{'}}e^{i\mathbf{q}^{''}.\mathbf{R}^{''}},
 \label{eq:8}
\end{eqnarray}
which depend on both the third-order interatomic force constants, $\phi^{\alpha\beta\gamma}\nonumber $, and eigenvectors of the phonons, $e_{\lambda}^{\alpha}(j)$. Here, $j$,
$j^{'}$, and $j^{''}$ are the atomic indices and $\alpha$, $\beta$, and $\gamma$ stand for the Cartesian components.

Phonons can be scattered by the boundaries of a finite size crystal. Phonon-boundary scattering is important, especially at low temperature and in nanostructures. $\tau_{\lambda}^{B}$ the relaxation time of the phonon-boundary scattering can be obtained as \cite{key-41,key-42,key-43}:
\begin{equation}
\frac{1}{\tau_{\lambda}^{B}}=\frac{\left|v_{\lambda}\right|}{L}\frac{1-P}{1+P},
\label{eq:9}
\end{equation}
where $L$ is the length of the system and $P$ is specularity parameter which ranges from zero for a diffusive scattering to unity for a specular scattering process. Therefore, $\tau_{\lambda}^{B}$ includes the finite size effect in the thermal conductivity calculations.

Another source of scattering is associated with the silicon isotopes. Silicon has several isotopes among which, $^{28}$Si is the most abundant (92.23\%) and $^{29}$Si (4.67\%) and $^{30}$Si (3.1\%) are stable too. One needs to account for the isotope scattering in order to calculate the thermal conductivity of naturally occurring silicene. The phonon-isotope scattering due to mass-difference of isotopes can be expressed as \cite{key-29,key-40,key-44,key-45}:

\begin{eqnarray}
\frac{1}{\tau_{\lambda}^{iso}}=\frac{\pi}{2N_{0}}\sum_{l}g_{l}\sum_{\lambda^{'}}\left|e_{\lambda}^{*}(l).e_{\lambda^{'}}(l)\right|^{2}\delta(\omega_{\lambda}-\omega_{\lambda^{'}}),
\label{eq:10}
\end{eqnarray}
where $g_{l}$ is the mass variance parameter of the $l$th atom and defined as
\begin{equation} 
g_{l}=\sum_{i}f_{il}(1-\frac{M_{il}}{\bar{M}_{l}})^2,
\label{eq:11}
\end{equation}
with $f_{il}$ being the concentration of $i$th isotope, $M_{il}$ is the mass of $i$th isotope and $\bar{M}_{l}$ is the average atomic mass of the $l$th atom. For silicene, because of its monatomic structure, we have $f_{il}=f_{i}$ and $M_{il}=M_{i}$.  

From the other side, real crystals naturally possess vacancy defects which affect the lattice thermal conductivity, significantly.
Using perturbation theory, Klemens studied the scattering of phonons from the static imperfections such as vacancy defects \cite{key-46,key-47}. Phonon-vacancy scattering is expressed in terms of the missing mass, missing linkages and change of the force constant between under coordinated atoms near the vacancy defects. The phonon relaxation time due to the missing mass and missing linkages can be determined from\cite{key-46,key-47,key-48}

\begin{equation}
\frac{1}{\tau_{\lambda}^{V_{1}}}=x(\frac{\triangle M}{M})^{2}\frac{\pi{\omega_{\lambda}^{2}g(\omega_{\lambda})}}{N_{0}},
\label{eq:12}
\end{equation}
where $x$ is the concentration of vacancy defects and $g(\omega_{\lambda})$ is the phonon density of states. Also, the force constant change contribution to the relaxation time is given by \cite{key-49,key-50}
\begin{equation}
\frac{1}{\tau_{\lambda}^{V_{2}}}=x(\frac{\delta k}{k})^{2}\frac{2\pi\omega_{\lambda}^{2}g(\omega_{\lambda})}{N_{0}},
\label{eq:13}
\end{equation}
with $k$ being the average stiffness. The change in force constant of bonds between under-coordinated atoms in the bond-order-length-strength notation can be expressed as \cite{key-48,key-51,key-52,key-53}
\begin{equation}
(\frac{\delta k}{k})=\left[\frac{1+\exp[(12-z)/(8z)]}{1+\exp[(13-z)/(8z-8)]}\right]^{-m+2}-1,
\label{eq:14}
\end{equation}
where $z$ is the coordination number and $m$ is a parameter that represents the nature of the bond. For Si the value of $m$ has been optimized to be 4.88 \cite{key-48,key-51,key-52,key-53}.

\section{\label{sec:level3}Results and discussion}

\subsection{\label{sec:level2}Intrinsic Thermal Conductivity}

\begin{figure*}[!t]\centering
	\includegraphics[scale=0.9]{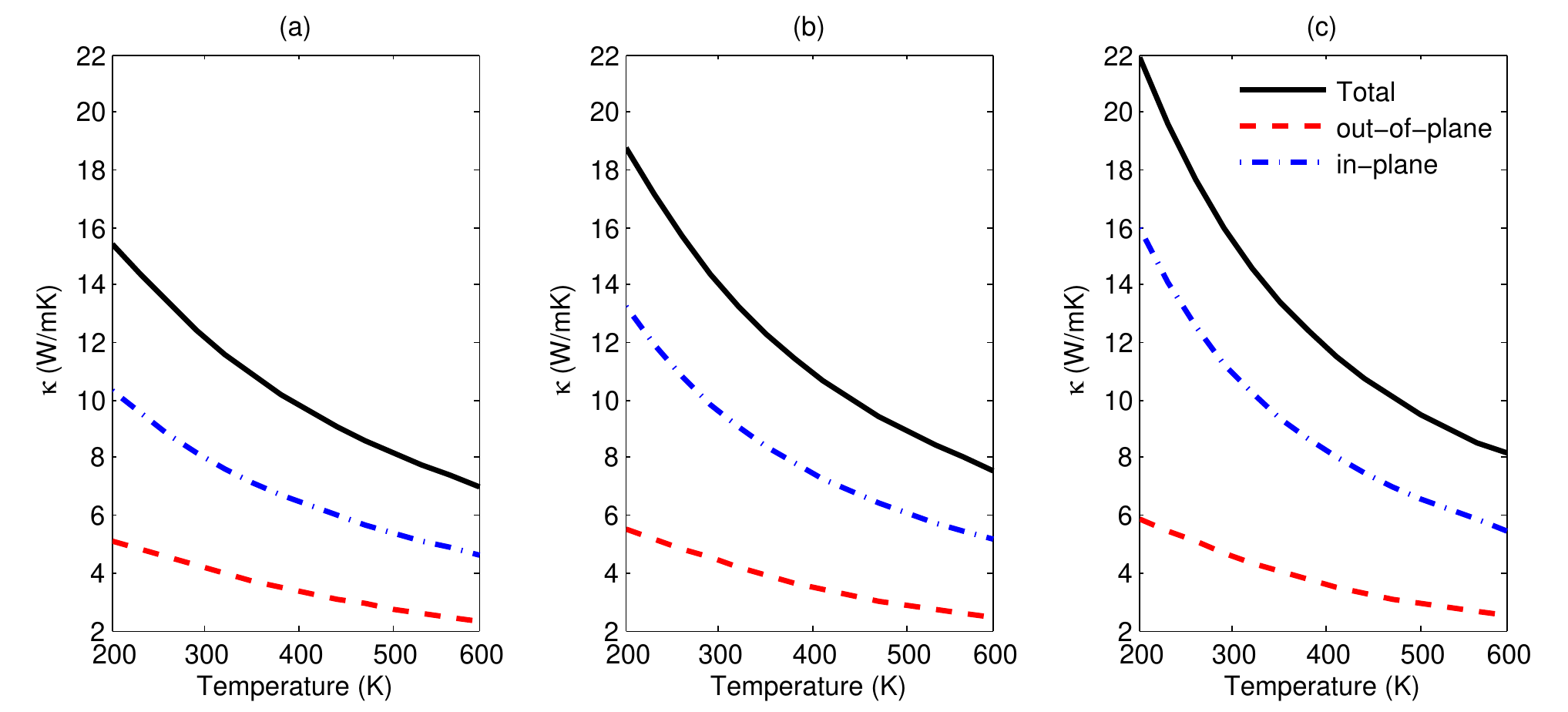} 
	\caption{Temperature dependence of the intrinsic thermal conductivity of silicene for different lengths with the same specularity parameter, $P=0$: (a) $L=100$ nm, (b) $L=300$ nm and (c) $L=3$ $\mu$m.} \label{fig3:sil}
\end{figure*}

To investigate the thermal transport in silicene, we first calculate the intrinsic thermal conductivity of the free-standing clean (pure and defectless) silicene by considering the phonon-phonon scattering with $P=0$ (a diffusive phonon-boundary scattering). Fig. \ref{fig3:sil} displays temperature dependence of the thermal conductivities obtained from the in-plane, out-of-plane and total modes contributions to the intrinsic thermal conductivity for three different lengths, $L=100$ nm, $300$ nm and $ 3\mu$m. It should be pointed out that the thermal conductivity of silicene with $L= 3 \mu$m is obtained around $15.5$ W/mK, while its value is about $150$ W/mK for the bulk silicon at room temperature. Similar decrease in thermal conductivity of the silicon nanowires and thin films has been experimentally reported \cite{key-19}. 

The presence of small buckling in silicene leads to a different contribution of phonon modes, compared to the planar structures; and as a result, an order of magnitude reduction in thermal conductivity is predicted. Moreover, strong scattering of the flexural modes from the in-plane phonon modes yields a low value for the thermal conductivity of silicene. We find that the in-plane acoustic modes are responsible for about 70\% of total intrinsic thermal conductivity of silicene at room temperature and play a major role in increasing  $\kappa$ with the sample length (see Fig. \ref{fig3:sil}), as it was pointed out earlier by Gu et al. \cite{key-21}.
In addition, it can be seen that increasing the temperature leads to a reduction in $\kappa$ which is expected for phonon-dominated systems in which the phonon-phonon scattering grows with temperature. Because of the buckled structure of silicene, the symmetry selection rule which applies to the case of graphene \cite{key-54}, does not apply here and therefore, the out-of-plane modes experience more scattering channels and as a result, their contribution to the thermal conductivity is smaller compared to the case of graphene  \cite{key-54}.
  
\subsection{\label{sec:level2} Effect of Phonon-Boundary Scattering}
 
\begin{figure*}[!t]\centering
 	\includegraphics[scale=0.9]{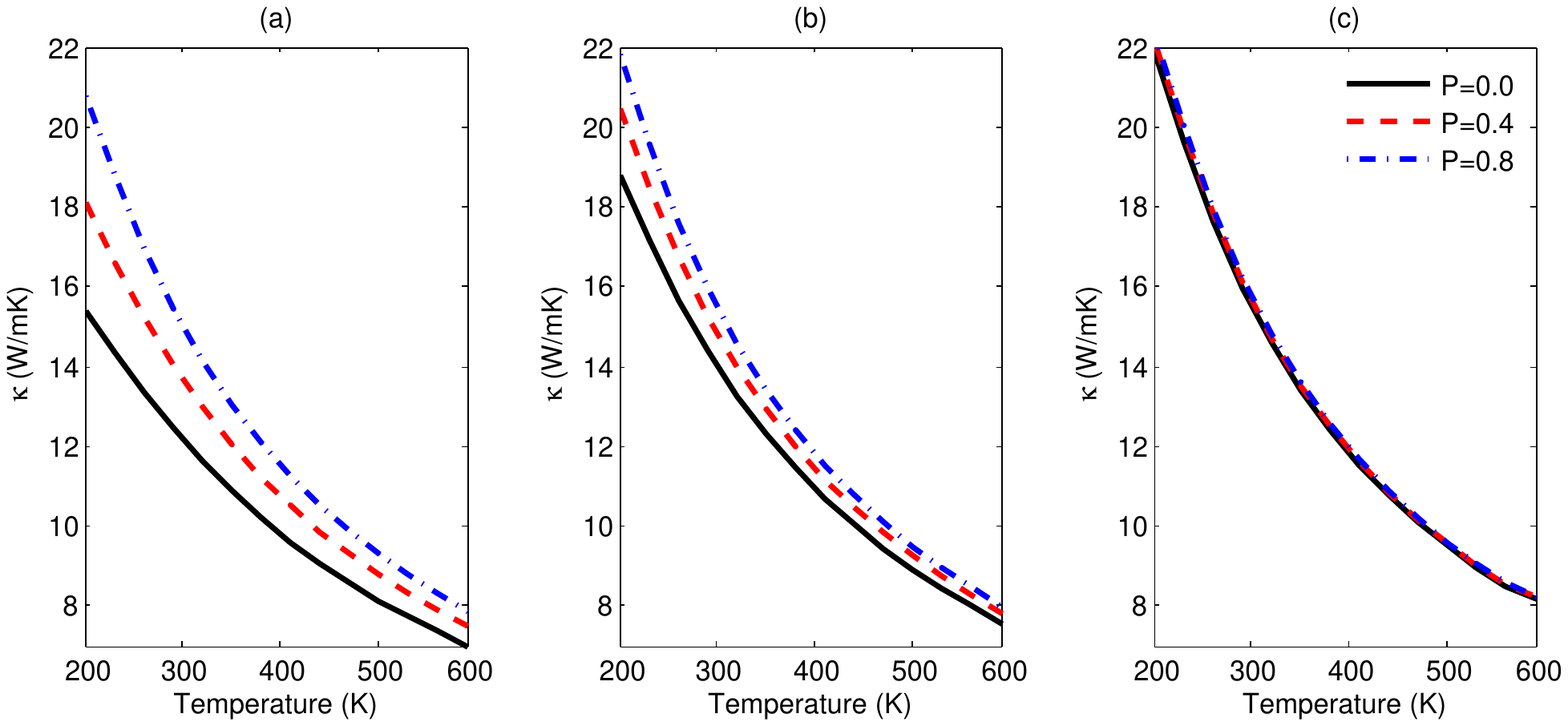} 
 	\caption{Temperature dependence of thermal conductivity of silicene for different specularity parameters, $P=0, 0.4$ and $0.8$ and different lengths:(a) $L=100$ nm, (b) $L=300$ nm and (c) $L=3$ $\mu$m.} 
 	\label{fig4:sil}
\end{figure*}

Fig. \ref{fig4:sil} shows the change of $\kappa$ with specularity parameter for three different sample sizes as a function of temperature. The thermal conductivity of pure and defectless silicene increases with growing the specularity parameter. When $P$ value varies from zero to unity, the nature of the phonon-boundary scattering turns from completely diffusive to specular and as a result, the thermal conductivity increases and the influence of sample size on the phononic transport becomes weaker.

Also, the change of the thermal conductivity of small-size silicene by the specularity parameter is more pronounced at low temperatures. In fact, increasing temperature enhances the strength of phonon-phonon scattering which is the dominant scattering mechanism and therefore, phonon-boundary scattering effect gets reduced. It is worth pointing out that the effect of phonon-boundary scattering is negligible for samples with large size (see Fig. \ref{fig4:sil}). Taking a look at Eq. (\ref{eq:9}), one learns that the relaxation time is proportional to the sample size which itself inversely contributes to the thermal conductivity, Eqs. (\ref{eq:1}) and (\ref{eq:3}). 

To complete this subsection, we should mention that because of a lower phonon group velocity in silicene, compared with graphene, the thermal conductivity of silicene is much less sensitive to the size effect (see Eq. (\ref{eq:9})).
 
\subsection{Effect of Phonon-Isotope Scattering}
  
The room-temperature thermal conductivity of naturally occurring silicene and pure silicene for two sample sizes are listed in table \ref{tabI:sil}. As it is shown, the effect of isotopic scattering in silicene is much smaller than that in graphene \cite{key-35,key-55}. This discrepancy stems from the difference in the masses of Si and C atoms. According to Eqs. (\ref{eq:10}) and (\ref{eq:11}), the isotopic scattering rate depends upon $(1-M_{il}/{\bar{M}_{l}})^2$. A larger atomic mass of Si with respect to C, makes this ratio smaller, resulting in a weaker isotopic scattering effect on the thermal conductivity.
\begin{table}[!h]
	\protect\caption{Thermal conductivity of pure and naturally occurring silicene at T=300 K.}	
	\begin{tabular}{|ccccc|ccccc|ccc|}
		\hline 
		&  & Size &  &  &  &  & {\large{$\kappa$}}${}_\mathbf{Pure}$ &  &  &  & {\large{$\kappa$}}${}_\mathbf{Natural}$ & \tabularnewline
		\hline 
		&  & 0.3 $\mu$m &  &  &  &  & 14 W/mK &  &  &  & 13.7 W/mK & \tabularnewline
		\hline 
		&  & 3  $\mu$m &  &  &  &  & 15.5 W/mK &  &  &  & 15 W/mK & \tabularnewline
		\hline 
	\end{tabular}
	\label{tabI:sil}
\end{table}
\subsection{Effect of Phonon-Vacancy Defect Scattering}

\begin{figure*}[!t]\centering
	\includegraphics[scale=0.9]{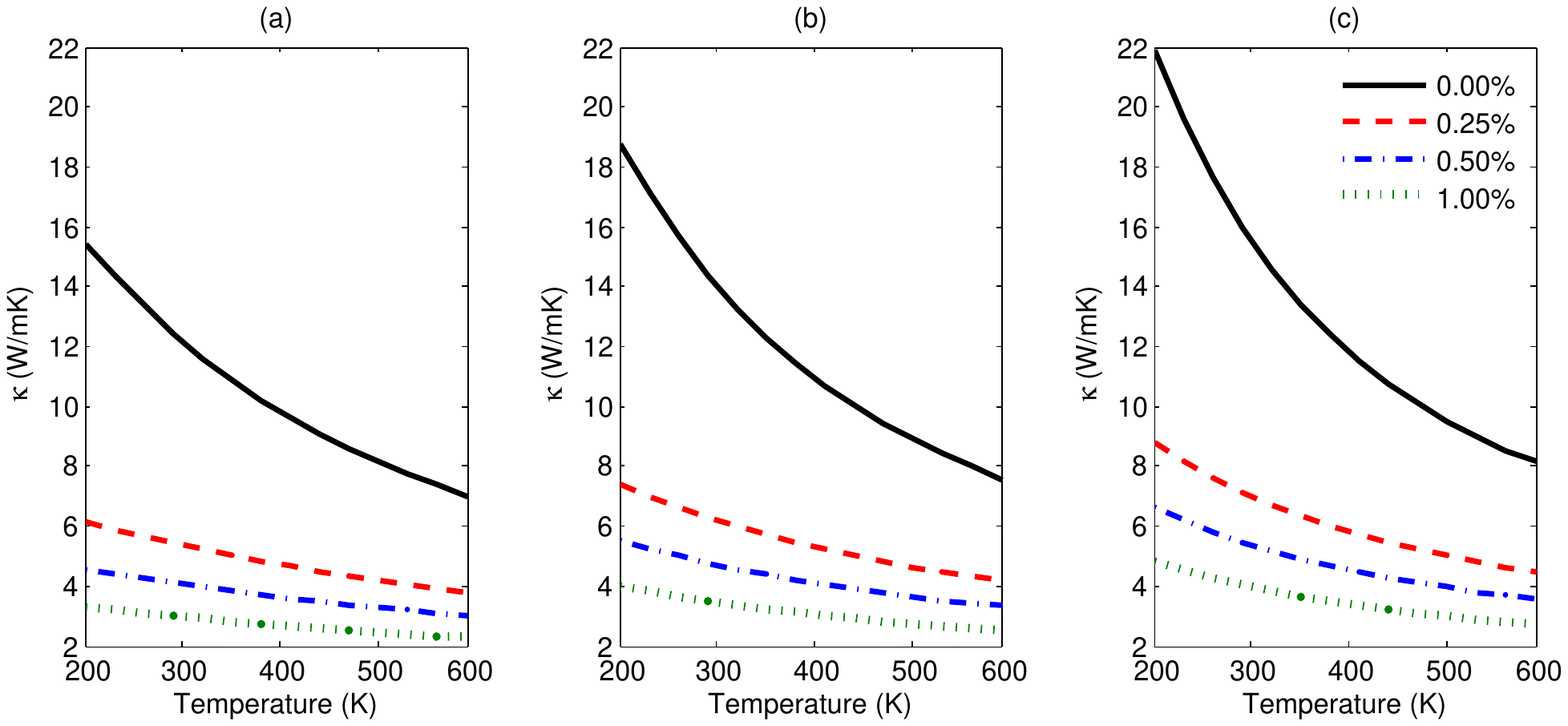} 
	\caption{Effect of single vacancy defects on the thermal conductivity of free-standing silicene for different lengths with $P=0$: (a) $L=100$ nm, (b) $L=300$ nm and (c) $L=3$ $\mu$m.}
	\label{fig5:sil}
\end{figure*}

In this subsection, we study the effect of single vacancy defects on the thermal conductivity of free-standing silicene. Fig. \ref{fig5:sil} shows the change of $\kappa$ with the single vacancy defect concentration as a function of temperature for three different sample lengths when the phonon-boundary scattering is considered to be purely diffusive ($P=0$).
\begin{figure}[!b]
	\includegraphics[width=9cm]{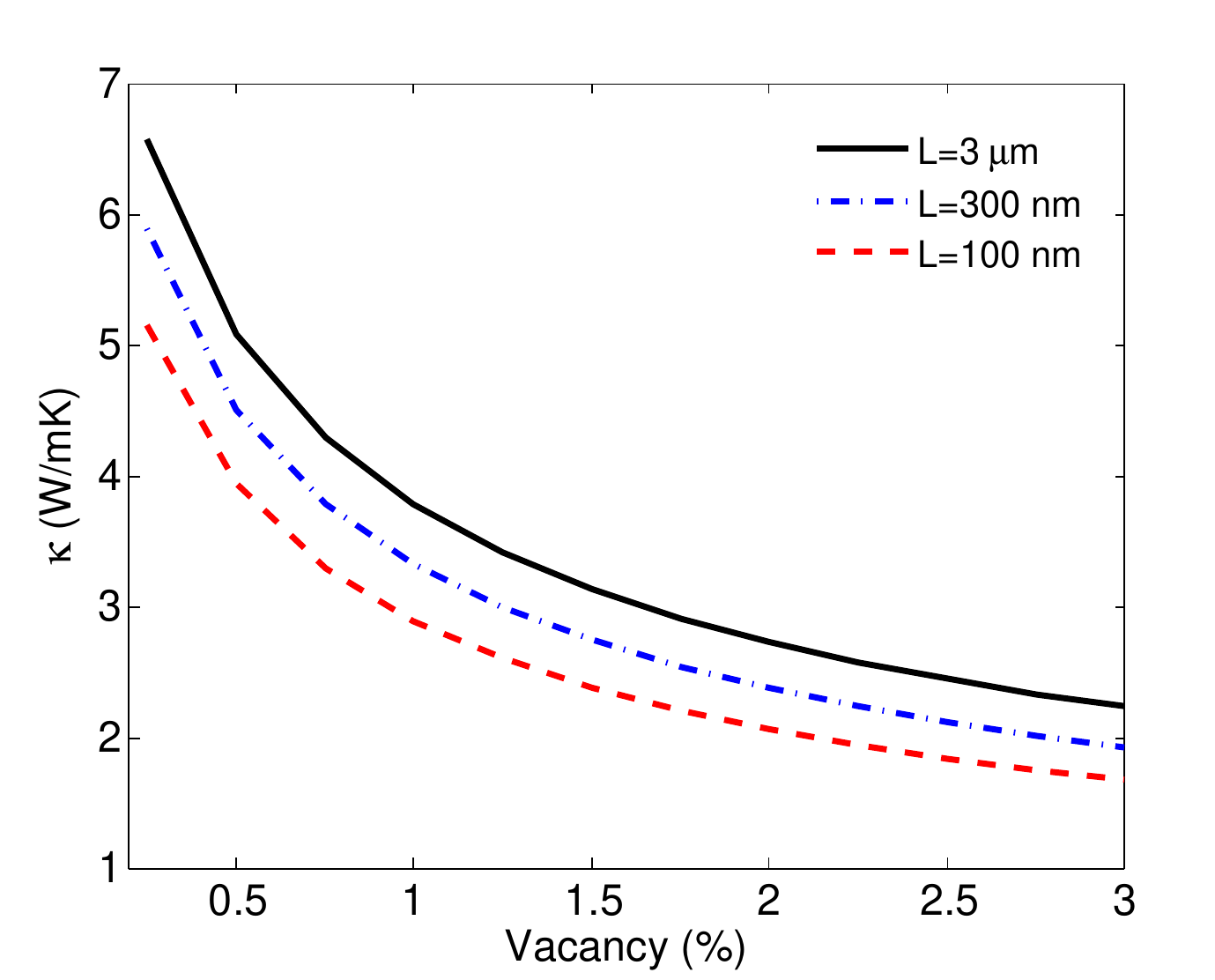} 
	\caption{Thermal conductivity of free-standing silicene versus vacancy concentration for three different lengths with $P=0$ at $T=300$ K.} \label{fig6:sil}
\end{figure}
The calculations presented in this figure suggest that the presence of a single vacancy defect significantly reduces the thermal conductivity values in silicene specially at low temperatures. By increasing temperature, the population of excited phonon modes increases and, as a result, the phonon-phonon scattering becomes dominant and makes the phonon-vacancy effect on the thermal conductivity weaker. Also, it can be observed that the thermal conductivity reduction due to the vacancy scattering is much stronger than the isotopic one (see Table \ref{tabI:sil}). A smaller coordination number in 2D materials compared to bulk materials, leads to a larger change in the force constants for those atoms which are close enough to the vacancy. Therefore, unlike the isotopic substitution effect, the phonon-vacancy defect scattering has a dominant impact on the thermal conductivity in silicene.
\begin{figure}[!b]
    \includegraphics[width=9cm]{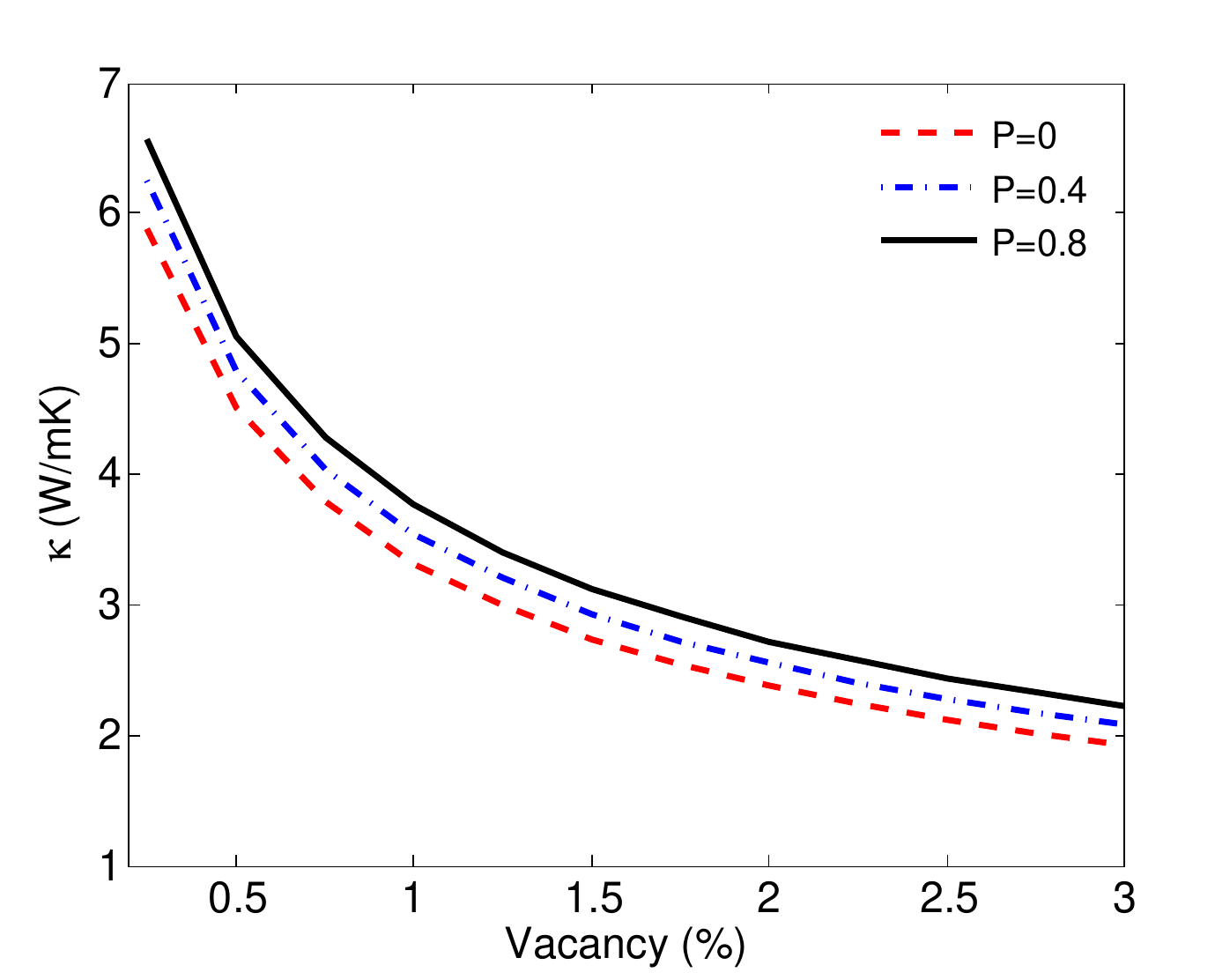} 
    \caption{Thermal conductivity of free-standing silicene versus vacancy concentration for different specularity parameters with $L=300$ nm at $T=300$ K.} \label{fig7:sil}
\end{figure}
\newline \indent To get clearer insight, we focus on the influence of the single vacancy defects concentration at a fixed (room) temperature. Fig. \ref{fig6:sil} depicts the changes of thermal conductivity of silicene versus single vacancy concentration for three different lengths at $T=300$ K. According to this figure, while $\kappa$ decreases remarkably with increasing vacancy percentage, this reduction is almost independent of the size of the sample. 
\newline\indent Our calculations show that even at the smallest single vacancy concentration studied here, 0.25 \%, {i.e.} the case in which only one of the every 400 atoms of Si is removed, the reduction in the thermal conductivity is substantial ($\approx 58\%$). Furthermore, our results predict that the significant reduction in the thermal conductivity of silicene happens for the concentrations less than 2\% and beyond that it slows down, gradually.
From Eqs. (\ref{eq:12}) and (\ref{eq:13}) it is obvious that the relaxation times due to the missing mass and linkages and change in the force constant are inversely proportional to the concentration of single vacancy defects. Therefore, an enhancement in the single vacancy concentration leads to a reduction in the relaxation time as well as its contribution to the thermal conductivity.
  
Also, variation of the thermal conductivity in silicene as a function of single vacancy concentration for three different specularity parameters with $L=3$ nm at room temperature is illustrated in Fig. \ref{fig7:sil}.        
One can see the reduction of $\kappa$ with increasing the vacancy percentage is not affected by the boundary scattering, similar to the result of Fig. \ref{fig6:sil}. This indicates that the effect of single vacancy defects in comparison to the effects of sample length and specularity parameter is dominant in silicene. 
\begin{figure}[!t]
	\includegraphics[width=9.5cm]{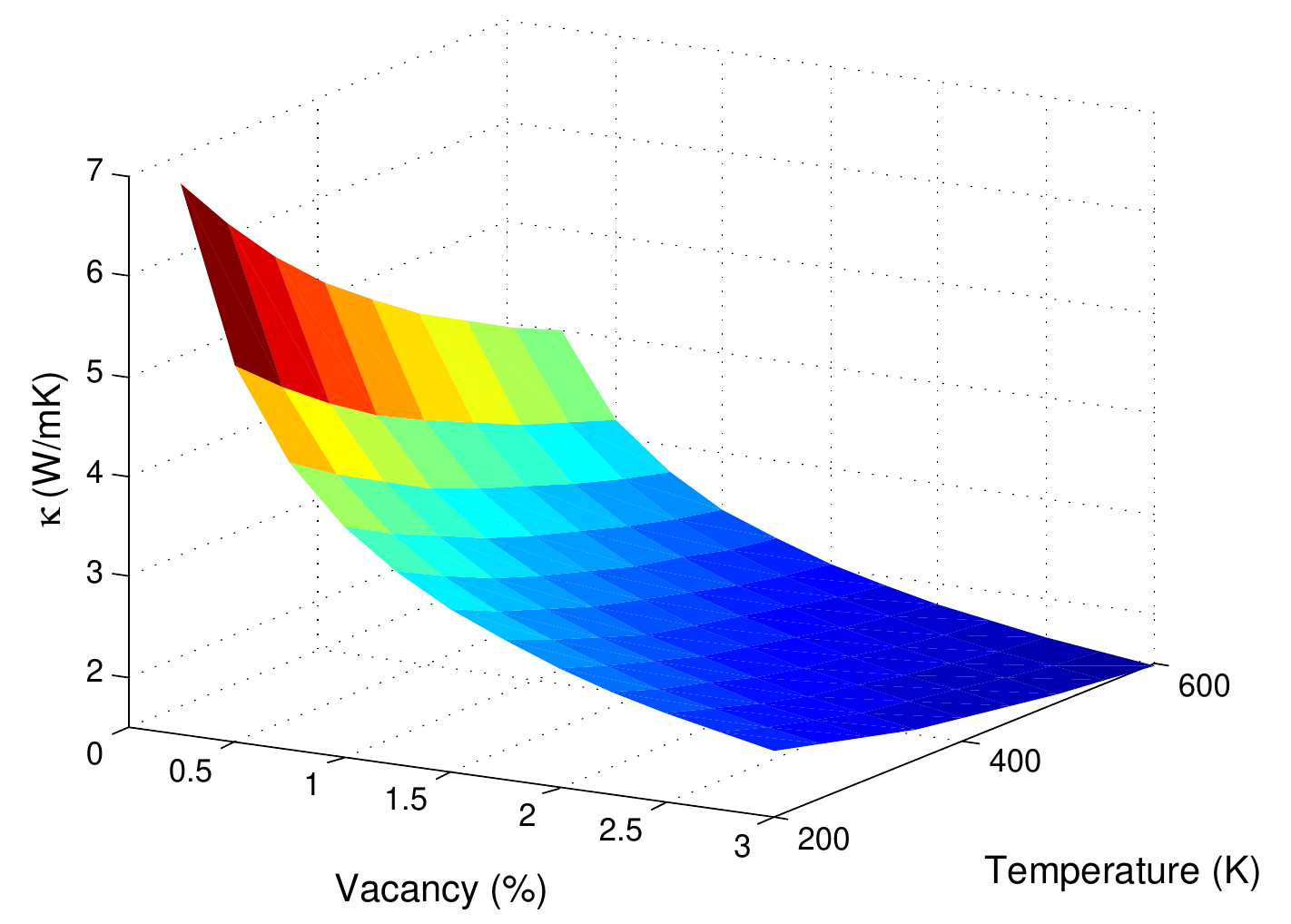} 
	\caption{Thermal conductivity of free-standing silicene as a function of single vacancy defect concentration and temperature with $L=100$ nm and $P=0$.}
	\label{fig8:sil}
\end{figure} 
In Fig. \ref{fig8:sil}, the changes of lattice thermal conductivity with the vacancy defect percentage and temperature are shown, simultaneously. As we can see, the effect of vacancy defects on the thermal conductivity of silicene is stronger than that of temperature.
    
\subsection{In-plane and Out-of-plane Phonons Contributions to the Thermal Conductivity in Presence of Different Scattering Mechanisms}
\begin{figure}[!t]
	\includegraphics[width=9cm]{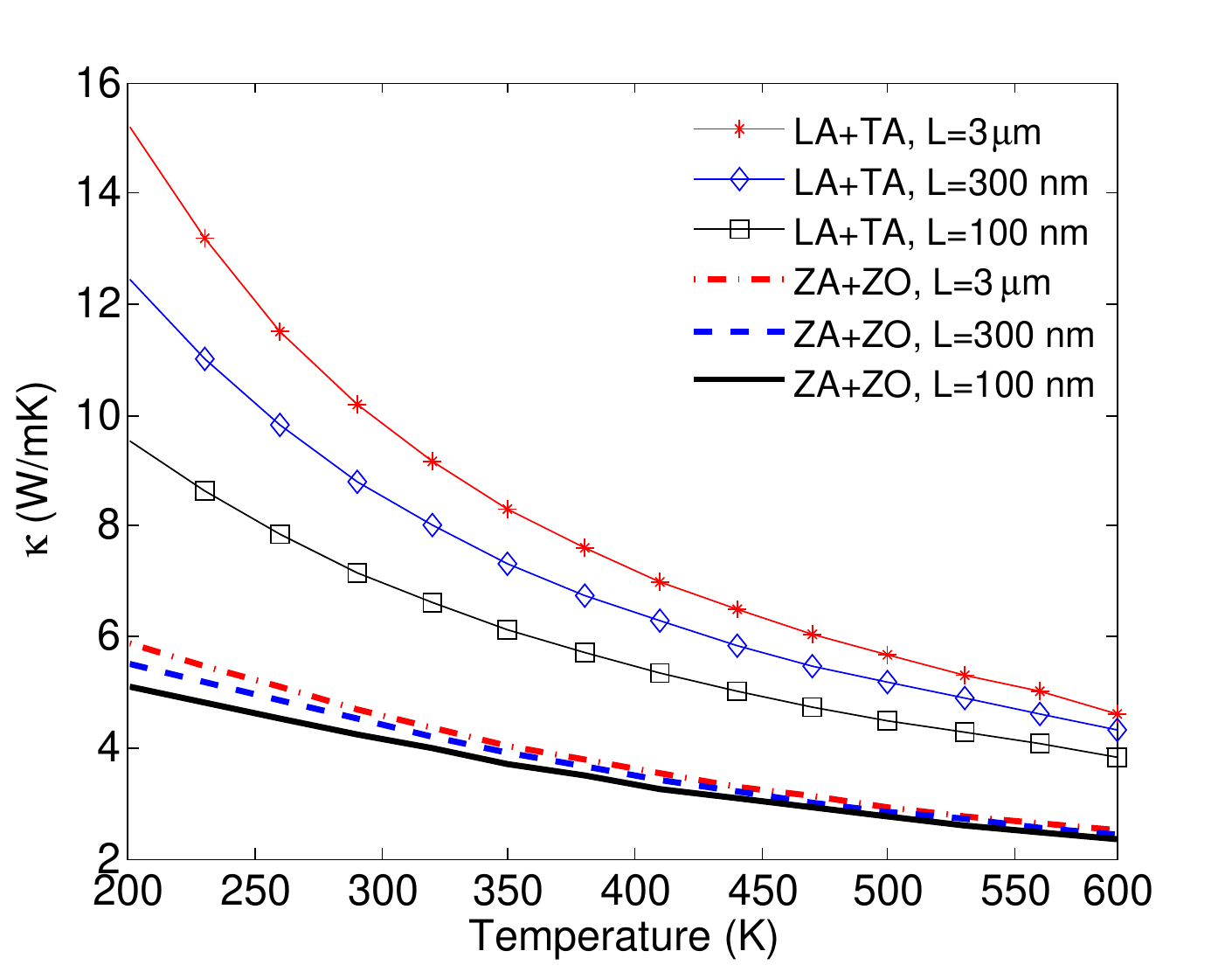} 
	\caption{Contributions to the thermal conductivity from the in-plane and out-of-plane phonon modes as a function of temperature for three different sample sizes with $P=0$.} \label{fig9:sil}
\end{figure}

In order to achieve deeper understanding of the phononic transport in silicene, we obtain the separate contributions from the in-plane and out-of-plane phonon branches to the thermal conductivity when another type of scattering mechanism exists. Fig. \ref{fig9:sil} illustrates the results of calculations for three different lengths of clean silicene with $P=0$. Because of small contributions of LO and TO modes at room temperature, these branches may be safely neglected. On the other hand, the dominant contribution to the thermal conductivity of silicene comes from the in-plane acoustic branches, which is about 70\% of the thermal conductivity for $L=3$ $\mu$m at $T=300$ K. As temperature increases, the contribution of optical branches grows and, for instance at $T=600$ K, the contribution of in-plane acoustic modes decreases to about 57\%.     
As mentioned earlier, due to the buckled structure of silicene in comparison to the other planar 2D structures, the symmetry selection rule does not apply \cite{key-21,key-54,key-55}. Therefore, the out-of-plane branches have more scattering channels than those in the planar structures. As a result, the contributions of the out-of-plane branches to the thermal conductivity are decreased. Moreover, increasing temperature strengthens the phonon-phonon scattering and weakens other effects like sample size effect (see Fig. \ref{fig9:sil}).

\begin{figure}[!t]
	\includegraphics[width=9cm]{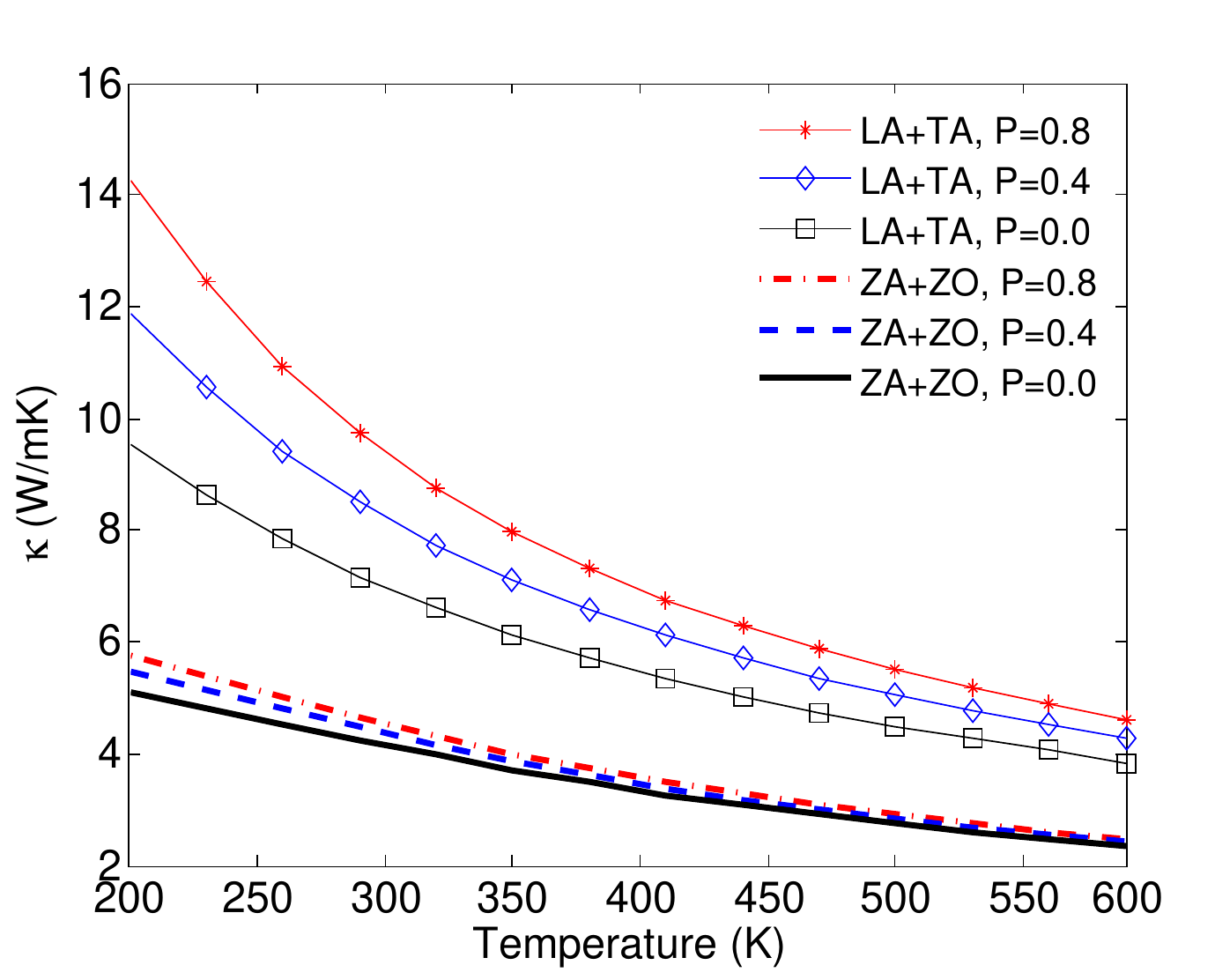} 
	\caption{Contributions to the thermal conductivity from the in-plane and out-of-plane phonon modes as a function of temperature for different specularity parameters with $L=100$ nm.} \label{fig10:sil}
\end{figure}

Also, the temperature dependence of the thermal conductivity associated to the in-plane and out-of-plane phonon branches of pure and defectless silicene with fixed sample size of $100$ nm for different specularity parameters is shown in Fig. \ref{fig10:sil}. As it is expected for 2D structures, the effect of phonon-boundary scattering on the contribution of the out-of-plane branches is very smaller than that in the case of in-plane branches. At high temperatures, again, it can be observed that the phonon-phonon scattering mechanism dominants and diminishes the influence of the phonon-boundary scattering.

\begin{figure}[!t]
   	\includegraphics[width=9cm]{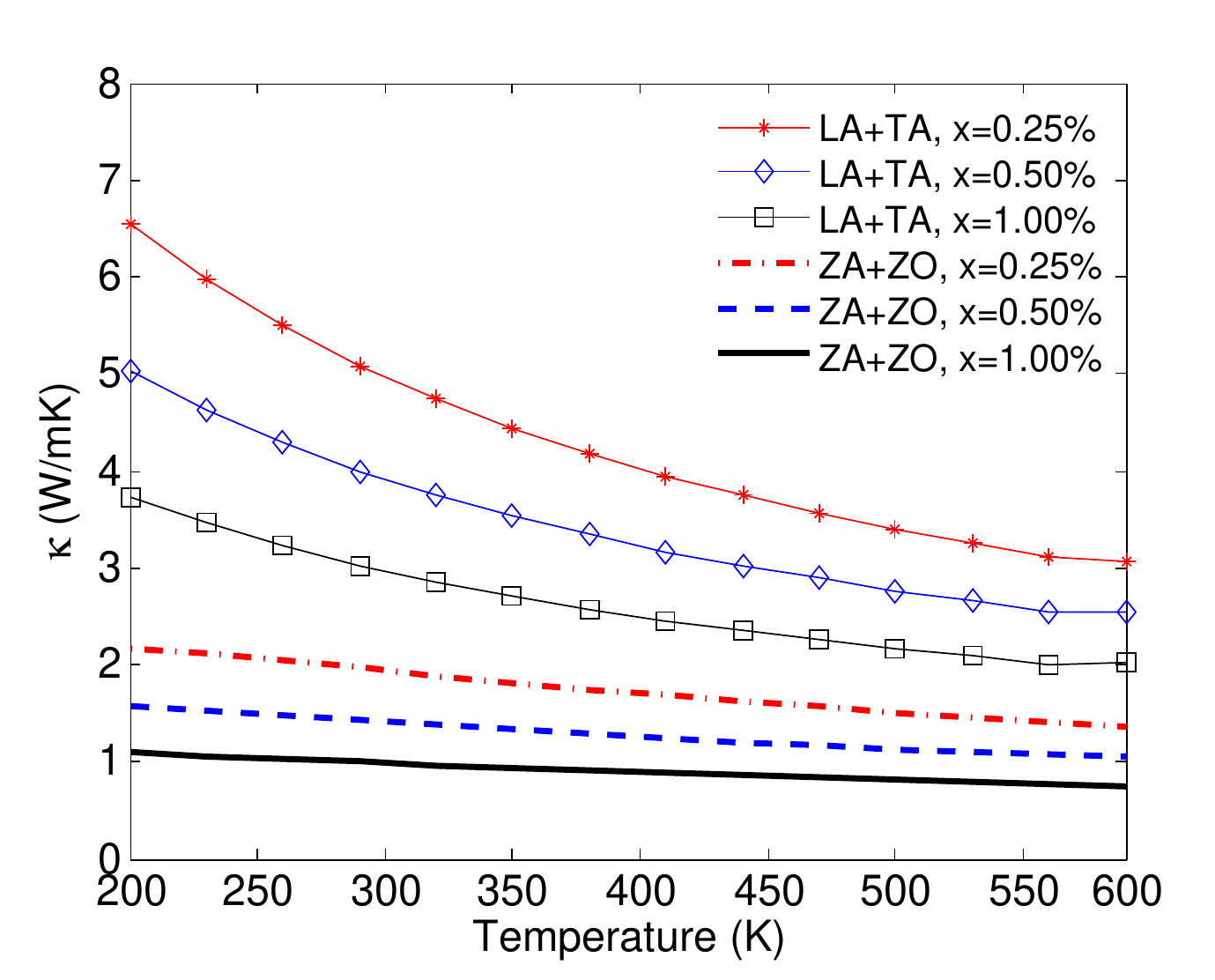}
   	\caption{Thermal conductivity of the in-plane and out-of-plane branches as a function of temperature for different concentrations of single vacancy defects with $L=3$ $\mu$m and $P=0$.} \label{fig11:sil}
\end{figure}

Finally, Fig. \ref{fig11:sil} displays the contributions from the in-plane and out-of-plane modes to the thermal conductivity as a function of temperature for three different concentrations of single vacancy defects for a sample size of $L=3$ $\mu$m with $P=0$.
Comparing this figure with Figs. \ref{fig9:sil} and \ref{fig10:sil}, we find that contrary to the previous cases, the contributions to the thermal conductivity from the out-of-plane phonon modes change considerably with the variation of vacancy defects concentration. However, the phononic transport due to these modes are less sensitive to the change of temperature. 
According to Fig. \ref{fig11:sil}, one may find out that the contributions from all phonon branches are strongly affected by the phonon-vacancy scattering, even at high temperatures, which highlights the important role of the vacancies in silicene for heat transfer applications. Also, we find that in the presence of vacancy defects, the contribution of in-plane acoustic modes to the thermal conductivity becomes larger with respect to the intrinsic case.
             
\section{\label{sec:level4}Conclusions}
In this paper, we have used the linearized PBTE within the RTA to calculate the lattice thermal conductivity of free-standing silicene under different scattering mechanisms including the phonon-phonon, phonon-boundary, phonon-isotope and phonon-vacancy defect. We have also studied the effects of sample size and different phonon modes on the phononic transport in silicene. We have employed the second- and third-order force constants, calculated with DFT, to obtain the phonon dispersion and phonon-phonon interactions, respectively. Our calculations have shown that the suppression of thermal conductivity by vacancies is quite remarkable so as even; removing only one of every 400 Si atoms leads to a substantial reduction in thermal conductivity. The present results have demonstrated that the impact of single vacancy defects on the heat transfer is much more pronounced at low temperatures. In addition, we have obtained that while the effect of isotopes on heat transfer is small, the role of phonon-boundary scattering is important in silicene samples with small size and no vacancies. Moreover, the contributions of the in-plane and out-of-plane branches to the thermal conductivity of silicene have been investigated. It has turned out that the contribution from the in-plane acoustic modes is dominant (about 70\% at room temperature) and grows as the vacancy population increases. Finally, we have found that, although the out-of-plane modes contribution to the thermal conductivity changes slowly with the sample size and specularity parameter, it depends clearly on the vacancy concentration. This work presents new insights into the heat conductivity of silicene which can become useful in nanoelectronic engineering and applications.

\section*{ACKNOWLEDGMENTS}

We would like to thank Dr. Ronggui Yang for providing us the third-order force constants of silicene through \url{http://spot.colorado.edu/~yangr/NUTS/NUTS_Pub.html}. It is also our pleasure to thank Dr. Lucas Lindsay for sharing some data on graphene with us.

\end{document}